\documentclass[conference]{IEEEtran}
\IEEEoverridecommandlockouts
\usepackage{cite, balance}
\usepackage{amsmath,amssymb,amsfonts}
\usepackage{enumitem}
\usepackage{graphicx}
\usepackage{textcomp}
\usepackage{caption}
\usepackage{subcaption}
\usepackage[utf8]{inputenc}
\usepackage{url}
\usepackage{xcolor}
\usepackage{color}
\usepackage[linesnumbered,ruled,vlined]{algorithm2e}
\def\BibTeX{{\rm B\kern-.05em{\sc i\kern-.025em b}\kern-.08em
    T\kern-.1667em\lower.7ex\hbox{E}\kern-.125emX}}

\newcommand{\pgm}[1]{{\textcolor{black}{#1}}}

\begin{document}

\title{Routing Heterogeneous Traffic \\in \pgm{Delay-Tolerant} Satellite Networks}

\author{Pablo G. Madoery,~\IEEEmembership{Member,~IEEE,} G{\"{u}}ne{\c{s}}~Karabulut~Kurt,~\IEEEmembership{Senior~Member,~IEEE}, \\Halim~Yanikomeroglu,~\IEEEmembership{Fellow,~IEEE}, Peng Hu,~\IEEEmembership{Senior~Member,~IEEE}\\ Khaled Ahmed, Stéphane Martel, Guillaume Lamontagne

\thanks{P. G. Madoery, G. Karabulut Kurt, and H. Yanikomeroglu are with the Non-Terrestrial Networks (NTN) Lab, Department of Systems and Computer Engineering, Carleton University, Ottawa, Canada, e-mail: \{pablomadoery, halim\}@sce.carleton.ca.}
\thanks{G. Karabulut Kurt is also with the Poly-Grames Research Center, Department of Electrical Engineering,  Polytechnique Montr\'eal, Montr\'eal, Canada, e-mail: gunes.kurt@polymtl.ca.}
\thanks{P. Hu is with the Digital Technologies Research Center, National Research Council Canada, Canada, email: Peng.Hu@nrc-cnrc.gc.ca.}
\thanks{K. Ahmed, S. Martel, and G. Lamontagne are with Satellite Systems, MDA, Canada.}
}

\maketitle
\IEEEpubidadjcol

\begin{abstract}
\pgm{Delay-tolerant} networking (DTN) offers a novel architecture that can be used to enhance store-carry-forward routing in satellite networks. Since these networks can take advantage of scheduled contact plans, distributed algorithms like the Contact Graph Routing (CGR) can be utilized to optimize data delivery performance. However, despite the numerous improvements made to CGR, there is a lack of proposals to prioritize traffic with distinct quality of service (QoS) requirements. This study presents adaptations to CGR to improve QoS-compliant delivery ratio when transmitting traffic with different latency constraints, along with an integer linear programming optimization model that serves as a performance upper bound. The extensive results obtained by simulating different scenarios show that the proposed algorithms can effectively improve the delivery ratio and energy efficiency while meeting latency constraints.

\end{abstract}

\begin{IEEEkeywords}
Routing, \pgm{delay-tolerant} networks, 
satellite constellations, QoS, contact graph routing, congestion avoidance.
\end{IEEEkeywords}

\section{Introduction}

The satellite industry has experienced unprecedented advances in microelectronics, commercial off-the-shelf solutions, and rocket launch platforms that have reduced the cost of manufacturing and deploying satellites \cite{9982444}. The total number of active satellites orbiting the Earth has increased from 1,459 in 2016 to 4,852 in 2022 and is planned to reach 100,000 satellites by 2030. In addition, the possibility of using inter-satellite links and on-board processing capabilities turns these satellite constellations into shared resource networks capable of providing ubiquitous sensing, navigation, positioning, and communication services.

The provision of this great diversity of services, with a heterogeneous network structure, reveals the need to use innovative communication protocols that can cope with variable delays and losses, as well as link disruptions caused by the high mobility of satellites with respect to terrestrial users and other satellites. 

In this regard, \pgm{delay-tolerant} networking (DTN) has been identified as a novel approach that considers delays and disruptions not as failures but as characteristics of this type of networks~\cite{RFC4838}. Therefore, the proposed architecture allows addressing these challenges from the design of the communication protocols and without requiring an excessive network infrastructure, such as the current satellite mega-constellations that aim to completely avoid such interruptions or delays. The fundamental part of the architecture is the Bundle Protocol (BP) that exists at a layer above the transport layers of the network. BP proposes the use of persistent storage on each DTN node to store-carry-and-forward data packets as transmission opportunities become available.

In the case of satellite networks, the subsequent communication episodes can be determined in advance based on orbital dynamics. These types of deterministic DTNs are referred to as scheduled DTNs, and can leverage a \textit{contact plan} that comprises the future network topology in order to optimize routing. The contact graph routing (CGR) algorithm, described in \cite{Fraire2021Routing}, is the most mature development for these networks and has been the focus of extensive contributions from the research community. These include the adaptation of the Dijkstra shortest path algorithm to time-varying networks \cite{segui2011enhancing}, the prevention of routing loops and consideration of multiple destinations \cite{birrane2012analysis, caini2021schedule}, source routing extensions \cite{edward2011improving}, and 
the application of overbooking management \cite{bezirgiannidis2014contact, bezirgiannidis2016analysis}. In addition to this, other improvements include congestion mitigation techniques \cite{madoery2018congestion}, route table management strategies and the incorporation of Yen’s algorithm \cite{fraire2018route}. Furthermore, the
incorporation of opportunistic \cite{burleigh2016toward} and probabilistic contacts \cite{RAVERTA2021102663}, the design of new schemes to enhance the scalability \cite{madoery2018managing}, a spanning-tree formulation to compute routes to several destinations \cite{de2019efficient}, and a partial queue information sharing \cite{dhara2019cgr}, as well as many other contributions.

\pgm{Additionally, recent works have begun to incorporate the required quality of service (QoS) as an important attribute to be taken into account in routing \cite{8766128,s21196356,guo2023online}. In  particular, \cite{8766128} proposes a centralized approach that involves the creation of a storage time aggregated graph (STAG) to represent the time-varying topology, followed by the assignment of mission’s data to specific routes in the STAG. Each mission has different latency requirements and the objective is to maximize the number of missions delivered to the destination. Using a centralized scheme has some advantages such as the potential to leverage global knowledge of both topology and traffic to derive more efficient solutions. On the other hand, centralized schemes suffer from some disadvantages such as overhead of communication with a central entity, increased complexity, lower scalability, single point of failure problem, reduced capability to react to failures, etc.}

Nevertheless, although the community has shown considerable interest, no substantial proposals have been made to adapt a distributed algorihtm such as CGR to prioritize traffic with different QoS requirements. Hence, this paper has a twofold contribution: first, we suggest modifications to the route selection process to prioritize traffic based on the required QoS. Second, we model the problem using a \pgm{integer} linear programming optimization model that provides a performance upper bound. \pgm{This paper is an extended version of \cite{9926911} and includes additional contributions such as an improved description of the \textit{CGR-Hops} routing scheme, a new multi-objective routing scheme called \textit{CGR-MO}, and new performance metrics for evaluation. In addition, we have extensively evaluated a realistic Walker-Delta constellation scenario and our numerical results demonstrate improvements in delivery ratio and energy efficiency while meeting maximum delay requirements.}

\pgm{It is worth noting that the scope of these contributions is framed within sparse and mid-size satellite networks that are well suited to be modeled with the DTN paradigm, which targets networks with high propagation delays and/or disruptions \cite{RFC4838}. Most of the time, the nodes of these networks do not have  end-to-end connectivity through multi-hop routes. Therefore, they must be able to buffer traffic when no communication is possible, perhaps for long periods of time, and send the traffic when communication becomes available. Although mega-constellations of satellites could take advantage of the proposed routing schemes to forward delay tolerant traffic, this case is beyond the scope of this paper.}

The remainder of the paper is organized as follows. In Section~\ref{sec:routing schemes}, we describe the processes involved in routing in scheduled satellite networks, including a description of the CGR algorithm and proposed adaptations to account for QoS requirements. Additionally, we provide an optimal traffic flow model together with an execution example comparing the behavior of the routing schemes in a specific scenario. Next, in Section~\ref{sec:evaluation} we describe the simulation environment, two different simulated scenarios, the performance metrics used and the results obtained.
Finally, we describe future lines of research in Section \ref{sec:futurework}, and conclude the work in Section \ref{sec:conclusion}.

\section{Routing Schemes}
\label{sec:routing schemes}

\subsection{Routing in scheduled satellite networks}
\label{sec:routing scheduled}
Four typical processes are involved in transmitting traffic from origins to destinations using scheduled communications within \pgm{delay-tolerant} satellite networks:

\begin{itemize}
    \item \textbf{Planning}: the determination of contact plans relies on a central entity, such as a ground station or mission control center, which estimates future communication episodes. \pgm{Each episode of communication between 2 nodes is known as a contact and has a start time, an end time, and an associated data rate}. To complete this task, it is necessary to consider the physical arrangement and orientation of nodes over time, along with their communication system configuration, which includes aspects such as antenna, transponder, modulation and transmission power. The combination of orbital propagators and communication models yields a contact plan that can be adjusted to minimize energy consumption or eliminate conflicting contacts \pgm{\cite{7105656, 8477064, 9750956}}. Once the contact plan is generated, it is disseminated to the satellites, which can then execute distributed routing processes when traffic is generated or received by different nodes within the network. \pgm{Although the design of the contact plan may have an important effect on the QoS obtained, this paper focuses on routing and therefore considers scenarios where the contact plan is trivial or is already defined.}
    \item \textbf{Routing}: in contrast to stable networks like the Internet, where routes are established as a series of nodes, each satellite in a delay-tolerant satellite network builds a routing table using the contact plan as an input. The routes in these networks are constructed as a sequence of temporary contacts and have associated time periods during which they are valid (they become invalid when one of their contacts terminates). Various attributes, such as estimated delivery time, number of hops/contacts, and remaining capacity, can be calculated for these routes.
    \item \textbf{Forwarding}: If a satellite generates or receives traffic, it consults its routing table, filters out routes that are no longer valid, and selects the one that optimizes a metric, typically delivery time. Using the information linked to the selected route, the satellite places the traffic in a queue that corresponds to a specific neighboring node (also called the next hop). 
    \item \textbf{Transmission}: Finally, when contacts are established between satellites, the enqueued traffic is transmitted in a specific order.
\end{itemize}

\subsection{Contact Graph Routing (CGR)}
\label{sec:cgr}
As detailed in \cite{Fraire2021Routing}, CGR is a routing algorithm that operates in a distributed fashion on several satellites. The algorithm's first task is to build a routing table (routing process), which is followed by the selection of routes for enqueuing traffic (forwarding process).

To construct the routing table, the approach utilizes the contact plan as input and constructs a contact graph structure, where vertices correspond to contacts and arcs represent data retention events in a satellite. Although this static structure may not be immediately intuitive, it simplifies the execution of route search algorithms in a topology that varies over time. The search process utilizes the contact graph structure and is an adapted version of Dijkstra's algorithm that explores the different contacts (nodes in the contact graph) and selects those that determine the optimal route in terms of earliest delivery time to the destination. To compare it with other schemes, we will refer to this scheme as \textit{CGR-DelTime}.
Moreover, CGR computes not just one route to a destination but instead the $K$ best routes using Lawler's modification of Yen's algorithm. This is due to the fact that routes have a deadline and limited capacity.

Subsequently, in the second stage, during the forwarding process, CGR begins with a packet and the routing table constructed in the preceding stage. It filters out routes that are either no longer viable or have reached their capacity and then chooses the most suitable route based on the earliest delivery time metric. Following this, the process is reiterated for every new packet that needs forwarding, with annotations indicating the capacity used from the route.

It is important to note that CGR considers the time-to-live (\textsf{TTL}) of a packet during the process of filtering routes. It disregards routes that have a projected delivery time later than the required time. However, selecting a route based solely on estimated delivery time could result in a situation where the route has numerous contacts (hops) that need to transmit traffic generated by other nodes. This may cause congestion issues where downstream nodes cannot honor the selected route, and the traffic fails to reach the destination within the required timeframe. 

It is, therefore, useful to make a clear distinction between the estimated delivery time and the effective delivery time. The estimated delivery time is a metric computed \textit{locally} at each node, and in the case of CGR it only takes into account topological aspects, i.e., what is the starting time of the last contact of a route. On the other hand, the effective delivery time is the time in which a packet is effectively delivered to its destination. This time may be later than the estimated delivery time due to traffic generated by intermediate nodes or unforeseen \textit{global} network conditions.
In general, the greater the number of hops in the chosen route, the greater the difference between the effective delivery time and the estimated delivery time. 

\subsection{Proposed Contact Graph Routing Adaptations}
\label{sec:cgradaptation}

\subsubsection{CGR-Hops}
 
To address the above-mentioned problems, this work considers a variation of CGR in the forwarding stage.
This new scheme called \textit{CGR-Hops} will still filter out those routes that do not deliver a packet in the required time, like the original CGR-DelTime scheme, but will then choose the best route as the one with the lowest number of hops. 

Therefore, CGR-Hops does not propose simply changing the delivery time metric by the number of hops in all the processes of CGR. The first filtering stage that \textit{CGR-DelTime} performs to let through only those routes whose estimated delivery time is less than the \textsf{TTL} is maintained. It is after this process that CGR-Hops chooses, among the routes that passed the filtering, the one with the lowest number of hops. This allows, on the one hand, to take advantage of the locally estimated delivery time of a route and on the other hand, not to choose routes with fewer hops but with late delivery times.

It is anticipated that this approach will take decisions that ease congestion by utilizing a smaller number of hops. This reduction in congestion is expected to free up capacity for traffic that requires earlier arrival.

\subsubsection{CGR-MO}
To explore the selection of routes with intermediate values of the number of hops and estimated delivery time, we propose a multi-objective adaptation of CGR called \textit{CGR-MO}. \textit{CGR-MO} maintains the same routing and route filtering stages as \textit{CGR-DelTime} and \textit{CGR-Hops}. The difference is during the forwarding stage and after route filtering. \textit{CGR-MO} calculates a metric $M$ associated with each of the computed routes and then chooses the route with the lowest value of $M$.
$M$ is defined according to Equation \ref{eq:M}, where $w$ is a weight that can vary in the range $[0,1]$, and where $norm\_hops$ and $norm\_delivery\_time$ are attributes of each route but normalized using the information present in the contact plan.
In particular, $norm\_hops$ is computed as the $hops\_number$ of the route divided by the total number of nodes in the network, while $norm\_delivery\_time$ is computed as the $estimated\_deli very\_time$ of a route divided by the contact plan time horizon.
Therefore, the values of $w$ closer to $1$ will give more importance to the number of hops, providing a behavior closer to \textit{CGR-Hops}, while values of $w$ closer to $0$ will give more importance to the delivery time, providing a behavior closer to \textit{CGR-DelTime}.

\begin{equation}
    M = w \times norm\_hops + (1-w) \times norm\_delivery\_time
\label{eq:M}
\end{equation}

\subsection{Optimal Traffic Flow \pgm{Integer} Linear Programming Model}
\label{sec:lpmodel}

To assess various routing strategies, it is beneficial to count with an optimization model. Although it may not be feasible to apply this model in practical networks due to computational complexity, it can still be utilized as an upper limit of performance to evaluate the proposed techniques in smaller networks.

The literature refers to the challenge of optimally routing multiple traffic as ``multi-commodity flow" \cite{Hu:1963}. Several studies, such as \cite{Jain:2004, Alonso03,fraire2016-tacp, 8766128}, have presented \pgm{integer} linear programming models that can solve this problem in networks with intermittent and scheduled communications, specifically in the context of DTN networks. In this paper, we introduce a modified version of these models, customized to our particular requirements, and extended to accommodate traffic with varying quality of service in terms of latency.

\subsubsection{Coefficients and Decision Variables}

The network topology is divided into $K$ states, where each state $k_q \in K$ represents the network during a specific time interval $[t_{q-1}, t_{q}]$. Therefore, the network is composed of $k_1,...,k_q,...k_f$ states, which correspond to the time intervals $[t_{0},t_{1}],...,[t_{q-1},t_{q}],...,[t_{f-1},t_{f}]$. The graph $G_{k_q}$ represents the network in each state $k_q$, where the vertices $v \in V$ are nodes, and the arcs $e \in E_{k_q}$ are communication opportunities (contacts) between nodes in $k_q$. The notation $c_{e}$ represents the capacity, measured in the number of packets, that can be transmitted from the source node to the destination node of an arc $e$ during one state $k_q$. Additionally, the number of packets that can be stored in the buffer of node $v$ is represented as $b_v$.

For a more concise notation, we will use $I^v$ to denote the set of arcs $e$ entering a node $v$, and $O^v$ to denote the set of arcs leaving it. Moreover, $D$ represents the set of all the traffic demands $d_{t_q, TTL}^{y,z}$ generated at time $t_q$ from node $y$ to node $z$ with a time-to-live of \textsf{TTL}. We assume that all traffic demands are generated at the beginning of a state, but if this is not the case, it is possible to add an intermediate state to account for such traffic generation.

Regarding the output variables, $X_{e}^{y,z}$ denotes the number of packets of traffic $y,z$ sent on arc $e$, while $B_{v, t_q}^{y,z}$ represents the number of packets of traffic $y,z$ stored at node $v$ at time $t_q$. Table \ref{Tab-LP} summarizes these model parameters.

\begin{table}[t]
\centering
\caption{Optimal Traffic Flow Model Parameters}
\label{Tab-LP}
\begin{tabular}{|c|c|}
\multicolumn{2}{c}{Input Coefficients}                                                                                                 \\ 
\hline
$t_q \in T$              & Timestamps                                                                                                  \\ 
\hline
$k_q =[t_q-1,t_q] \in K$ & States (Time intervals)                                                                                     \\ 
\hline
$v \in V$                & Nodes                                                                                                       \\ 
\hline
$e \in E_{k_q}$            & Arcs of a graph in state $k_q$                                                                              \\ 
\hline
$c_e$                    & Capacity of arc $e$                                                                                         \\ 
\hline
$b_v$                    & Capacity of buffer's node $v$                                                                               \\ 
\hline
$d_{t_q, TTL}^{y,z} \in D$        & \begin{tabular}[c]{@{}c@{}}Traffic from node $y$ to node $z$ \\originated at timestamp $t_q$ \\with time-to-live \textsf{TTL}\end{tabular}   \\ 
\hline
\multicolumn{1}{l}{}     & \multicolumn{1}{l}{}                                                                                        \\
\multicolumn{2}{c}{Output Variables}                                                                                                   \\ 
\hline
\pgm{$\{X_{k_q, e}^{y,z}\}$}            & \begin{tabular}[c]{@{}c@{}}\pgm{Traffic from node $y$ to node $z$}\\ \pgm{sent in state $k_q$ through arc $e$}\end{tabular}                 \\ 
\hline
$\{B_{t_q,v}^{y,z}\}$        & \begin{tabular}[c]{@{}c@{}}Buffer occupancy of node $v$ \\at timestamp $t_q$ by traffic $y,z$\end{tabular}  \\
\hline
\end{tabular}
\end{table}

\subsubsection{Objective Function and Constraints}

Taking into account the coefficients and decision variables defined, we link them by means of the objective function defined in \eqref{LPModel01}
subject to the constraints \eqref{LPModel02} to \eqref{LPModel07}.

Thus, an \pgm{integer linear programming (ILP)} model is formulated with inputs comprising a set of traffics ($d_{t_q,TTL}^{y,z}$) to be transmitted from sources ($y$) to destinations ($z$) through a time-varying topology with buffer capacities ($b_v$) and transmission capacities ($c_e$), and outputs consisting of optimal flows (\pgm{${X_{k_q, e}^{y,z}}$}) to be follow by the traffics.

The objective function described in equation \eqref{LPModel01} aims to minimize the sum of products $w(k_q)\times$\pgm{${X_{k_q, e}^{y,z}}$}, where $w(k_q)$ is a weighting function that assigns a higher weight to each state in increasing order. Therefore, minimizing this product prioritizes the delivery of traffic as soon as possible while minimizing the use of later arcs.

Several constraints are imposed in the model to ensure a consistent transmission of traffic. Equation \eqref{LPModel02} defines the buffer occupancy of each node ($v$) at each timestamp ($t_q$) using the previous state ($t_{q-1}$), the incoming ($e \in I^v$) and outgoing ($e \in O^v$) flows in the interval $[t_{q-1}, t_q]$, and the traffic generated at $t_q$ ($d_{t_q, TTL}^{y,z}$). This equation must hold for all $t_q,v,y,z$. Equation \eqref{LPModel03} limits the buffer occupancy ($B_{t_q,v}^{y,z}$) of each node to $b_v$, while equation \eqref{LPModel04} limits the traffic flow through each arc (\pgm{${X_{k_q, e}^{y,z}}$}) to $c_e$.

Moreover, constraint \eqref{LPModel05} establishes the buffers at the initial timestamp ($t_0$), and constraint \eqref{LPModel06} ensures that traffic generated in intermediate states is not sent in earlier states. This constraint also guarantees that the traffic that has a maximum latency constraint (\textsf{TTL}) arrives at its destination on time and remains there until the final state. Finally, equation \eqref{LPModel07} ensures that all traffics generated over time reside in their respective buffers at the destination nodes ($B_{t_f,v}^{y,z}$) at timestamp $t_f$. 

Table \ref{Tab-LP} summarizes these model parameters.

\begin{flalign}
	\label{LPModel01}
	\textrm{minimize:} \quad \sum\limits_{k_q \in K} \sum\limits_{e \in E_{k_q}} \sum\limits_{y \in V} \sum\limits_{z \in V} w(k_q) * X_{k_q, e}^{y,z} 
\end{flalign}
Subject to:
\begin{equation}
\label{LPModel02}
B_{t_{q},v}^{y,z} = \begin{cases}
B_{t_{q-1},v}^{y,z} + \pgm{\sum\limits_{e \in I^v} X_{k_q, e}^{y,z}} - \pgm{\sum\limits_{e \in O^v} X_{k_q, e}^{y,z}} + d_{t_q,TTL}^{y,z} & \small{\text{$y = v$}}\\
B_{t_{q-1},v}^{y,z} + \sum\limits_{e \in I^v} X_{k_q, e}^{y,z} - \sum\limits_{e \in O^v} X_{k_q, e}^{y,z}  & \small{\text{$y \neq v$}}
\end{cases}
\end{equation}
\begin{flalign}
\label{LPModel03}
\sum\limits_{y \in V} \sum\limits_{z \in V} B_{t_q,v}^{y,z} &<= b_{v} \quad \forall \; t_q,v\\
\label{LPModel04}
\sum\limits_{y \in V} \sum\limits_{z \in V} X_{k_q, e}^{y,z} &<= c_{e} \quad \forall \; k_q, e
\end{flalign}
\begin{equation}
\label{LPModel05}
B_{t_0,v}^{y,z} = \begin{cases}
d_{t0,TTL}^{y,z} & \text{if $y = v$}\\
\quad 0 & \text{if $y \neq v$}
\end{cases}
\quad \forall \; v,y,z
\end{equation}
\begin{equation}
\label{LPModel06}
B_{t_q,v}^{y,z} >= \begin{cases}
d_{t_q,TTL}^{y,z} & \text{if $y = v$}\\
\quad 0 & \text{if $y \neq v$}
\end{cases}
\quad \forall \; t_q>t_0, v,y,z
\end{equation}
\begin{equation}
\label{LPModel07}
B_{t_f,v}^{y,z} = \begin{cases}
\sum\limits_{t_q \in T} d_{t_q, TTL}^{y,z} & \text{if $z = v$}\\
\quad 0 & \text{if $z \neq v$}
\end{cases}
\quad \forall \; v,y,z
\end{equation}

\subsection{Execution Example}
\label{sec:example}

\begin{figure*}[!tbh]
\centering
\begin{subfigure}{0.31\textwidth}
    \centering
    \fbox{\includegraphics[width=\textwidth]{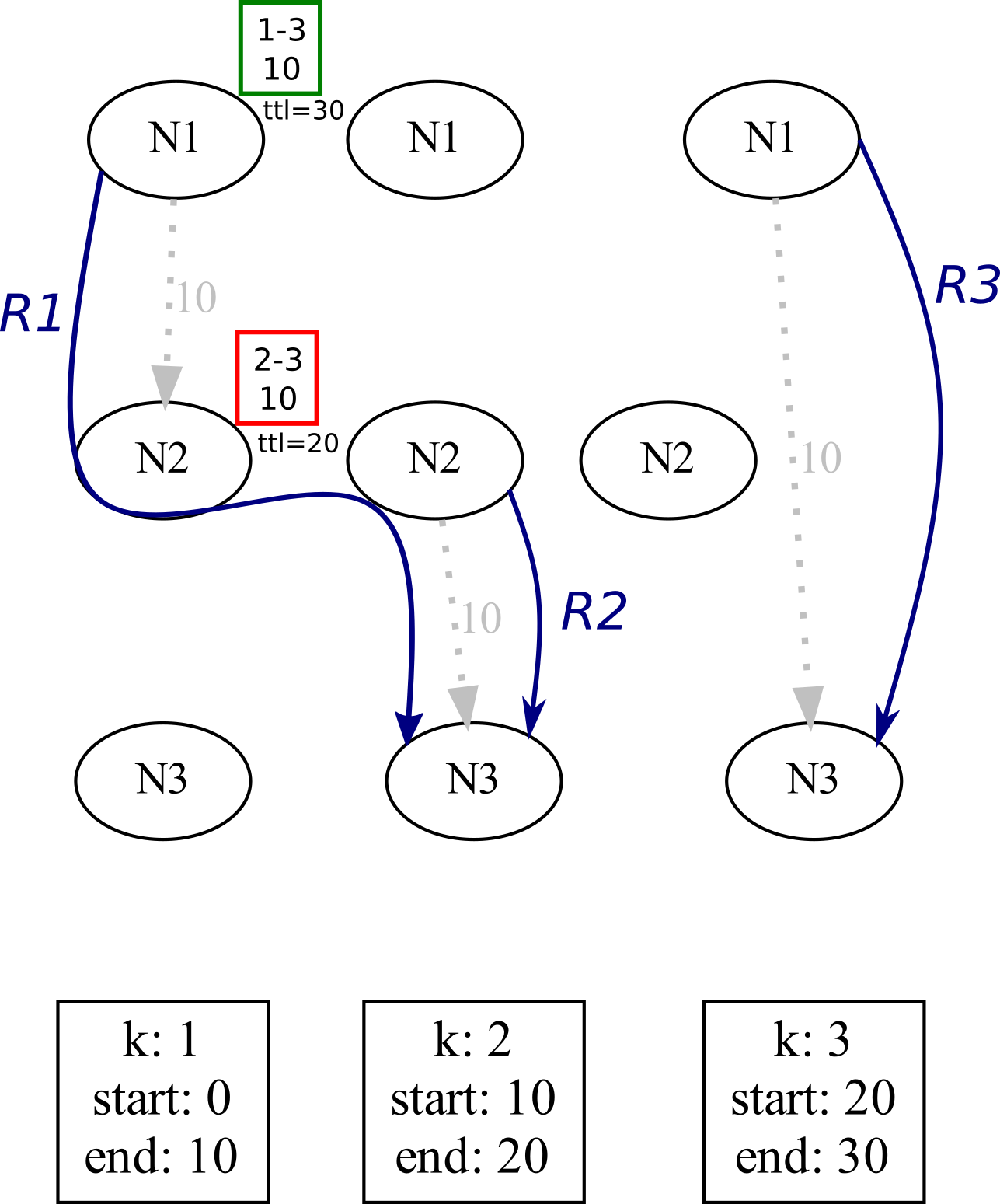}}
    \caption{Topology, traffic and routes}
    \label{fig:Topology}
\end{subfigure}
\hfill
\begin{subfigure}{0.274\textwidth}
    \centering
    \fbox{\includegraphics[width=\textwidth]{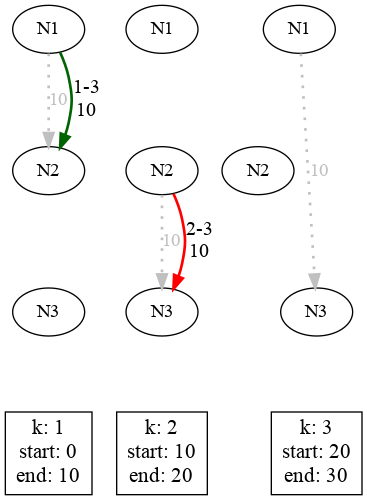}}
    \caption{\textit{CGR-DelTime} traffic flow}
    \label{fig:CGR-DelTime}
\end{subfigure}
\hfill
\begin{subfigure}{0.318\textwidth}
    \centering
    \fbox{\includegraphics[width=\textwidth]{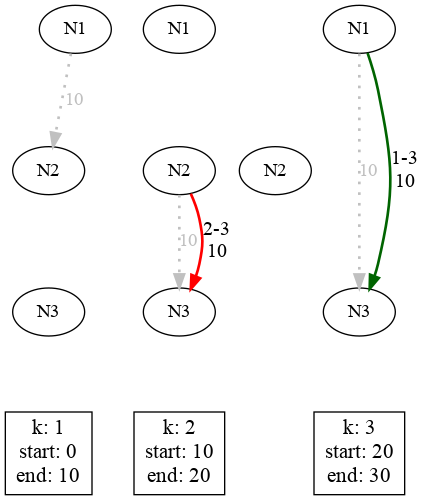}}
    \caption{\pgm{\textit{ILP model}} and \textit{CGR-Hops} traffic flow}
    \label{fig:Model}
\end{subfigure}
\caption{Example of execution of different routing schemes.}
\label{fig:example}
\end{figure*}

To gain an intuitive understanding of the routing schemes, we examine a basic example where three nodes ($N1$, $N2$, $N3$) each have 3 contacts in different states as illustrated in Figure \ref{fig:Topology}. The dotted lines show contacts with the capacity to send 10 packets each. Nodes $N1$ and $N2$ both produce 10 packets in state $k_1$ headed for $N3$ (contained in colored squares). $N1$'s traffic has a \textsf{TTL} of 30 seconds, so it must arrive at its destination by state $k_3$ or earlier, while $N2$'s traffic has a \textsf{TTL} of 20 seconds, requiring it to reach the destination by state $k_2$ or earlier. $N1$ can access routes $R1$ and $R3$ to reach $N3$, while $N2$ can use route $R2$, as depicted by solid blue lines.

The behavior of \textit{CGR-DelTime} is shown in Figure \ref{fig:CGR-DelTime}, where colored solid lines depict traffic transmissions. $N1$ calculates and sends traffic using route $R1$, but when the packets reach $N2$, they encounter an unexpected situation: $N2$ also has traffic destined for $N3$ and must use the entire capacity of the $N2$-$N3$ contact. As a result, congestion arises, preventing $N1$'s 10 packets from reaching $N3$ within the required time. Conversely, $N2$'s traffic reaches $N3$ on time by following route $R2$.

Finally, Figure \ref{fig:Model} displays the behavior obtained using both \textit{CGR-Hops} and the \textit{ILP model}. Although \textit{CGR-Hops} knows that route $R1$ delivers the packets sooner, it chooses $R3$, which not only delivers the packets on time (\textsf{TTL} = 30 s) but also uses fewer hops, resulting in congestion avoidance. This effect is achieved by both the \textit{ILP model} and \textit{CGR-Hops}, resulting in more QoS-compliant traffic flows, and enabling all traffic to reach its destination on time. However, the \textit{ILP model} requires global knowledge of the topology and traffic generated, whereas in this case \textit{CGR-Hops} can achieve similar outcomes without knowing the traffic generated by other network nodes. To assess the potential of this approach, we will evaluate and analyze whether this behavior persists in more complex scenarios.

\section{Performance evaluation}
\label{sec:evaluation}

\subsection{Simulation Environment}

To compare the performance of the routing schemes described in Section \ref{sec:routing schemes}, we have adapted and extended Dtnsim, a discrete event-driven simulator developed in the Omnet++ framework. DtnSim was originally presented in~\cite{Fraire:2017:DtnSim} and is available as open-source software in a public repository\footnote{The public DtnSim repository can be found at:~\url{https://bitbucket.org/lcd-unc-ar/dtnsim}}.

As shown in Figure \ref{fig:dtnsim}, each node (satellite or ground station) is modeled as an Omnet module with 3 internal sub-modules:

\begin{figure}[!b]
	\centering
	\includegraphics[width=0.482\textwidth]{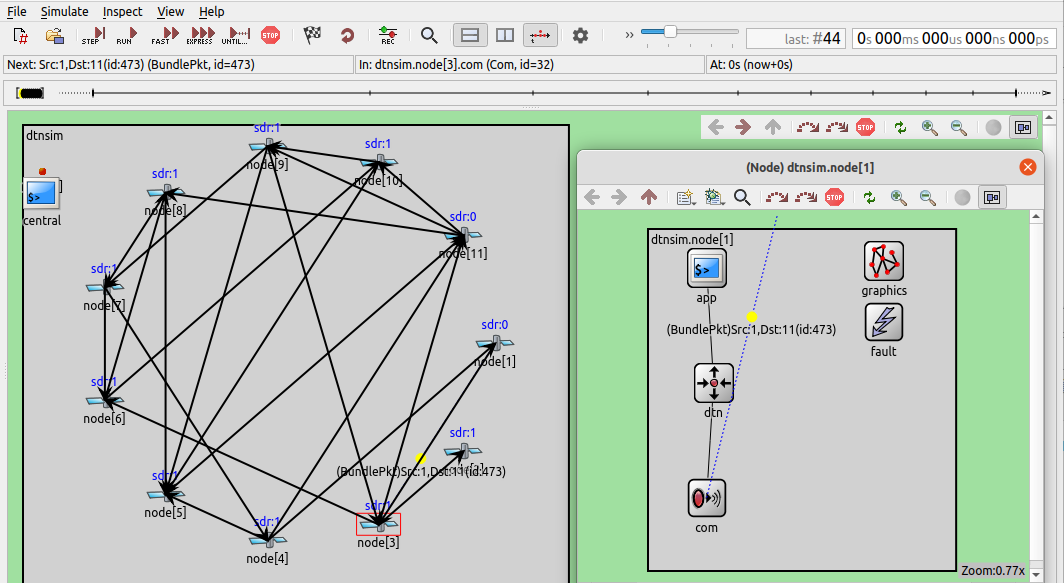}
	\caption{Dtnsim graphical interface with nodes exchanging packets (left), and sub-modules belonging to a node (right).}
	\label{fig:dtnsim}
\end{figure}

\begin{itemize}
    \item \textit{\textbf{app}}: in charge of generating and/or receiving traffic packets at a given time, with a source, a destination, a size in bytes, and a \textsf{TTL}. 
    This sub-module communicates directly with the $dtn$ sub-module.
    \item \textit{\textbf{dtn}}: in charge of storing, routing, and forwarding traffic packets. 
    This sub-module implements all the routing schemes described and communicates directly with the $app$ and $com$ sub-modules.
    \item \textit{\textbf{com}}: in charge of transmitting/receiving packets to/from other nodes. 
    This sub-module communicates directly with the $dtn$ sub-module and possibly with the com sub-module of other nodes.
\end{itemize}

In addition to these sub-modules, a \textit{\textbf{central}} module is responsible for reading a file containing the contact plan and distributing that information to all nodes of the network. Then, each node establishes contacts with its neighbors at the corresponding times. In this way, although the nodes remain visually static, the time-varying topology is captured and implemented through the input contact plan.

\begin{figure*}
	\centering
\includegraphics[width=1\textwidth]{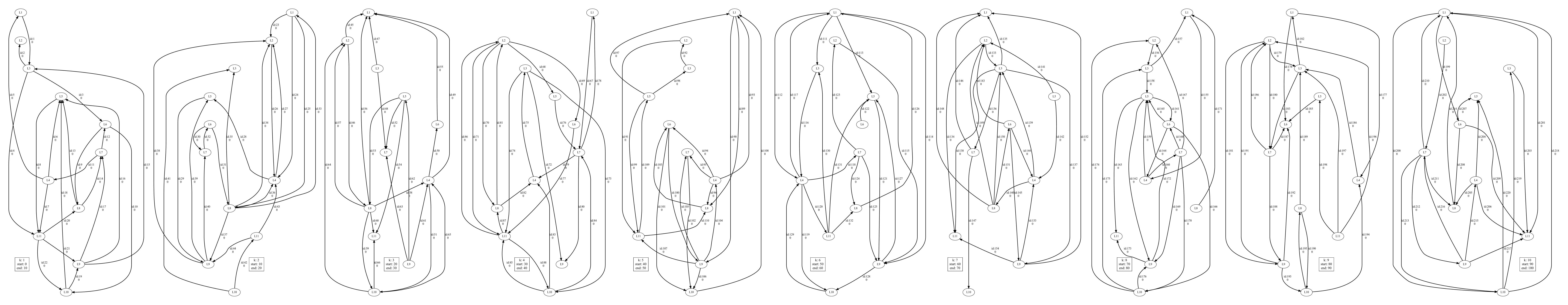}
	\caption{Example of a random network consisting of 10 states.}
	\label{fig:random_network}
\end{figure*}

\subsection{Simulation Scenarios}
\label{sec:scenarios}
As described below, we performed simulations of 2 types of scenarios: Random Networks and Walker-Delta Constellation:

\subsubsection{Random Networks}
We generated 100 random network topologies of 100s divided into 10 states of 10s. An example topology can be seen in Figure \ref{fig:random_network}. Each of these topologies model the time-varying connectivity among 11 nodes, with nodes 1-10 representing satellites and node 11 representing a ground station. 

The connectivity between nodes is governed by a contact density parameter $\delta$ that ranges from 0.0 to 1.0. A network with $\delta=1.0$ is fully connected with contacts present in all states, while $\delta=0$ implies no contacts exist. For our study, we set $\delta=0.2$. Increasing $\delta$ requires higher traffic generation to observe the same congestion and performance degradation. Each link has a capacity to send 10 traffic packets, and nodes have unlimited storage capacity.

To induce congestion, we used an all-to-one traffic pattern in which all 10 nodes generate variable traffic load directed to node 11. Nodes 1-5 generate packets without maximum latency requirements (\textsf{TTL} $\rightarrow$ $\infty$), allowing them to reach their destination in any state. In contrast, nodes 6-10 generate packets with a maximum latency requirement of 20 seconds, forcing them to arrive at their destination in either the first state ($k_1$) or the second state ($k_2$).

These parameters were chosen to ensure that heterogeneous traffic was routed, congestion was eventually provoked, and packets had a chance to arrive at their destination before the simulation ended.

An important clarification is that the random network generation consisted of two stages. A first stage in which 100 random networks were generated without any constraints, followed by a second stage in which only those networks were kept in which the \pgm{\textit{ILP model}} was able to find an optimal routing solution for the maximum traffic load generated. After these two stages, 25 random networks were obtained that comply with the described property.

\subsubsection{Walker-Delta Constellation}
This type of constellation was first proposed in \cite{walker1984satellite}, studied in several works such as \cite{Wood2003, Fraire2017-Hindawi, 8422917}, and is widely used by real deployments such as the Starlink constellation operated by SpaceX.  

\begin{figure}[!b]
	\centering
	\includegraphics[width=0.47\textwidth]{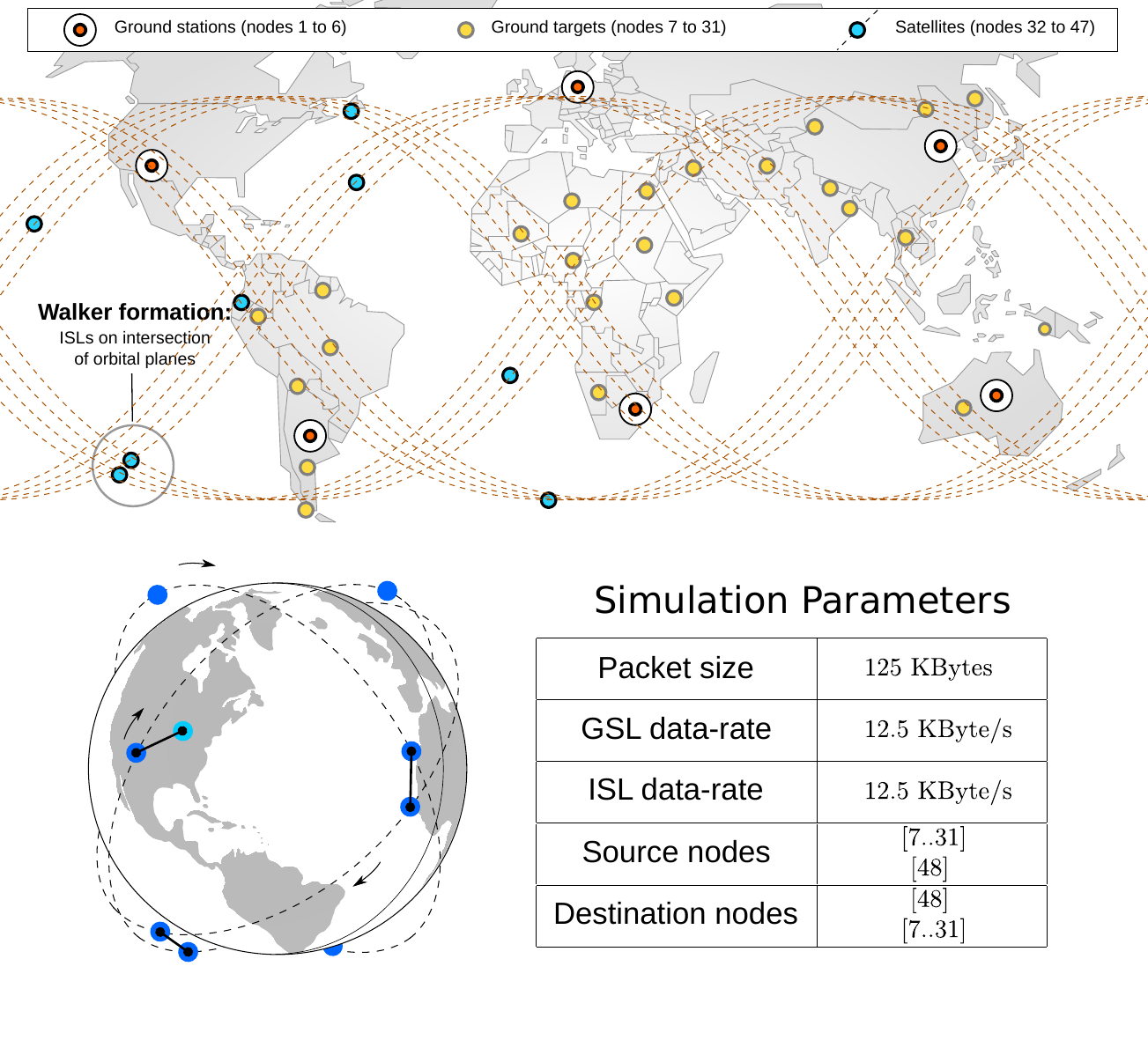}
	\caption{Walker-Delta constellation represented in 2 and 3 dimensions together with simulation parameters.  Figure adapted from \cite{Fraire2017-Hindawi}.}
	\label{fig:walker}
\end{figure}

For this work, we have used a sparser Walker-Delta constellation with the orbital and simulation parameters described in \cite{Fraire2017-Hindawi}, and represented in Figure \ref{fig:walker}. This particular type of network is one of the so-called Ring Road Networks (RRNs) \cite{9755273} and its purpose is to connect remote sites without internet access (\textit{cold spots}) with sites in urban regions with internet access (\textit{hot spots}) by means of small and inexpensive satellites that act as \textit{data mules} to exchange data between hot spots and cold spots. 

The constellation was propagated over a time horizon of 1 day and is composed of 4 orbital planes with 4 satellites per orbital plane. 
In addition, there are 25 ground targets (nodes 7 to 31), 6 ground stations (nodes 1 to 6), and a mission control center (MCC, node 48). The MCC is connected via the Internet to each ground station and satellites are capable of exchanging data through inter-satellite links (ISLs) that are established when they have line of sight and the distance between them is less than 1,000 km. Only 1 ISL needs to be established at the same time. Furthermore, satellites are capable of establishing links with ground stations (GSLs) when the same conditions are met. The resulting contact plan used for the simulations can be found in \url{https://upcn.eu/icc2017.html}.

With respect to traffic, each of the ground targets generates a variable traffic load destined to the MCC, while the MCC generates a variable traffic load destined to each of the ground targets. Half of the generated traffic packets have a \textsf{TTL} of 1 hour, while the other half has no \textsf{TTL} restriction (\textsf{TTL} $\rightarrow$ $\infty$). 

\begin{figure*}[!htb]
\begin{subfigure}{.5\textwidth}
  \centering
  \includegraphics[width=1.0\linewidth]{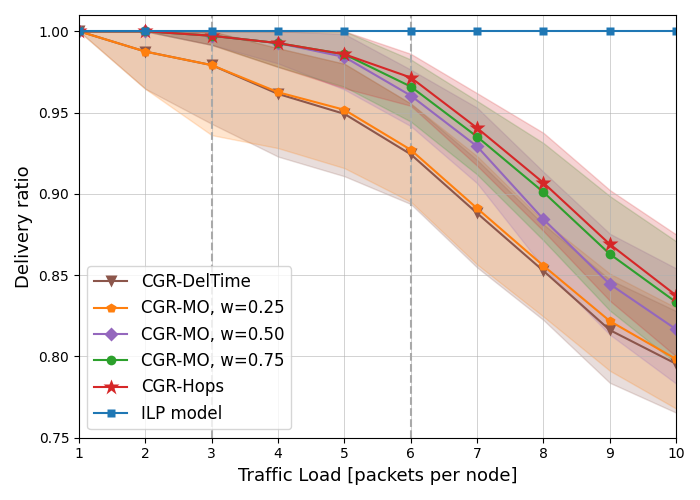} 
  \caption{Delivery ratio}
  \label{fig:random_delivery_ratio}
\end{subfigure}
\vspace{0.4cm}
\begin{subfigure}{.5\textwidth}
  \centering
  \includegraphics[width=1.0\linewidth]{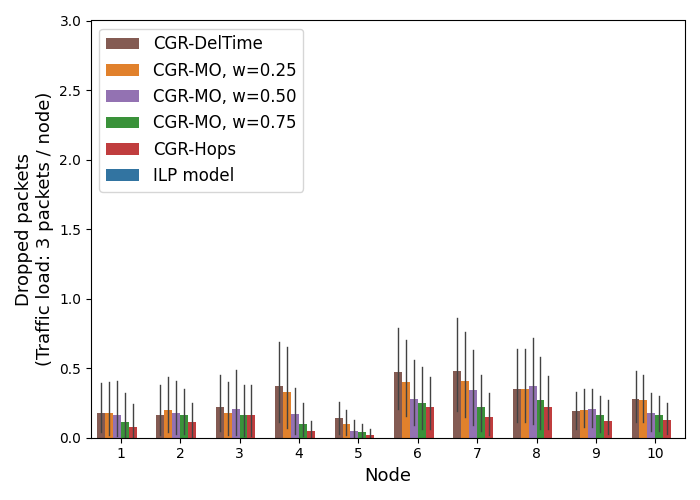}
  \caption{Dropped packets when traffic load is 3 packets per node}
  \label{fig:random_packets_dropped_3}
\end{subfigure}
\begin{subfigure}{.5\textwidth}
  \centering
  \includegraphics[width=1.0\linewidth]{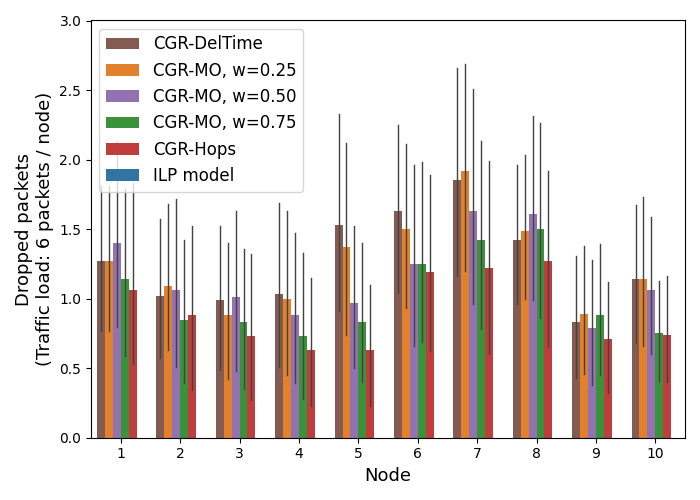} 
  \caption{Dropped packets when traffic load is 6 packets per node}
  \label{fig:random_packets_dropped_6}
\end{subfigure}
\vspace{0.4cm}
\begin{subfigure}{.5\textwidth}
  \centering
  \includegraphics[width=1.0\linewidth]{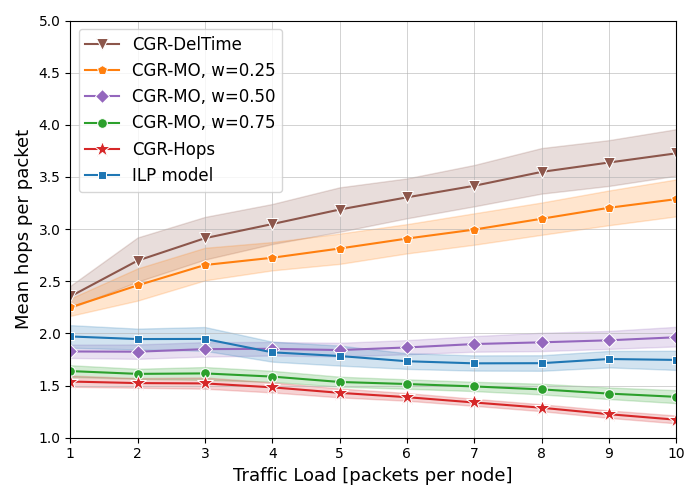}
  \caption{Mean hops per packet}
  \label{fig:random_mean_hops}
\end{subfigure}
\begin{subfigure}{.505\textwidth}
  \centering
  \includegraphics[width=1.0\linewidth]{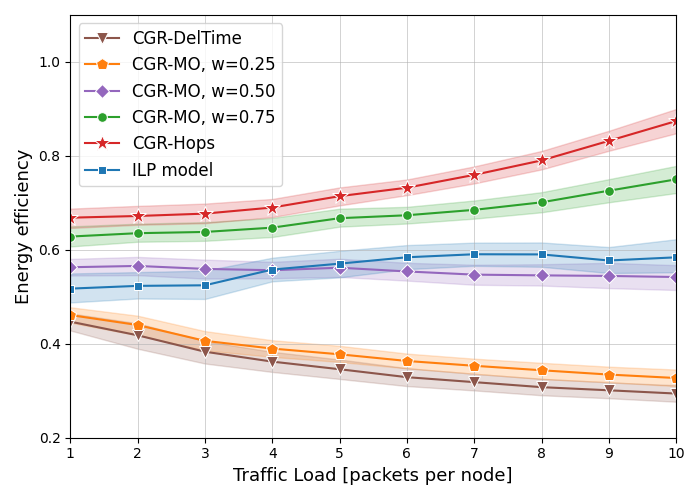}
  \caption{Energy efficiency}
  \label{fig:random_energy_efficiency}
\end{subfigure}
\begin{subfigure}{.505\textwidth}
  \centering
  \includegraphics[width=1.0\linewidth]{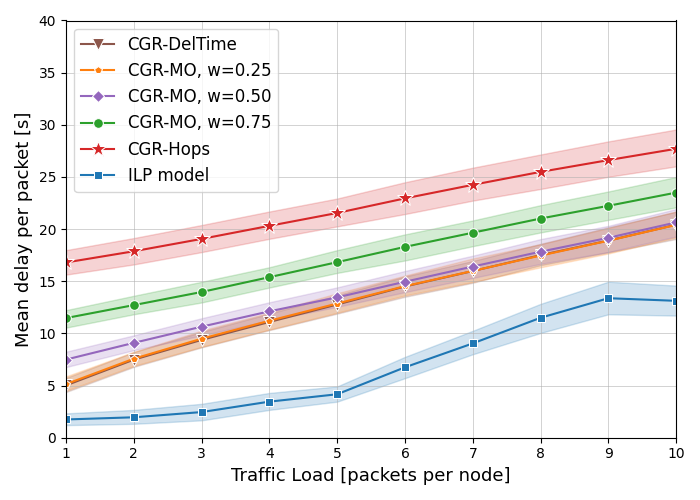}
  \caption{Mean delay per packet}
  \label{fig:random_mean_delay}
\end{subfigure}
\vspace{0.2cm}
\caption{Simulation results for random networks ($\delta=0.2$).}
\label{fig:random_results}
\end{figure*}

\subsection{Performance Metrics}

We compare the performance of \textit{CGR-DelTime}, \textit{CGR-Hops}, \textit{CGR-MO} with weights 0.25, 0.5, 0.75, and the \pgm{\textit{ILP model}} described in Section \ref{sec:routing schemes}. We will be examining the following metrics in our analysis:
\begin{itemize}
\item \textit{\textbf{Delivery ratio}}: the total number of generated packets divided by the total number of packets that arrive at their destination within their specified maximum latency.
\item \textit{\textbf{Dropped packets per node}}: the total number of packets dropped at each node. If a node's routing scheme cannot find a route that provides a chance of meeting the specified maximum latency, the packet is discarded. This increases the value of this metric and decreases the value of the delivery ratio metric.
\item \textit{\textbf{Mean hops per packet}}: the total number of transmissions of all packets divided by the total number of packets that arrive at their destination within their specified maximum latency.
\item \textit{\textbf{Energy efficiency}}: the total number of packets that arrive at their destination within their specified maximum latency divided by the total number of transmissions of all packets.
\item \textit{\textbf{Mean delay per packet}}: the average latency at which packets are delivered to their destination within their specified maximum latency. Packets that are dropped and do not reach their destination do not contribute to this metric.
\end{itemize}

\subsection{Simulation Results}

\subsubsection{Random Networks}
The performance metrics when different values of traffic load are generated at the source nodes are plotted in Figure \ref{fig:random_results}. The curves and bars reflect the average values and the shaded areas (and error bars) represent the 95\% confidence intervals.

Figure \ref{fig:random_delivery_ratio} shows that regardless of the traffic load, the \pgm{\textit{ILP model}} is able to deliver all generated packets to destination, that is, the delivery ratio is 1 and the dropped packets are 0. This is achieved thanks to the fact that the model has high computational complexity and is able to access global network information both in terms of topology and traffic. Furthermore, as mentioned in section \ref{sec:scenarios}, only cases where the \pgm{\textit{ILP model}} is able to find a feasible solution are considered in this analysis. On the other hand, the rest of the schemes have a delivery ratio that decreases as the traffic load increases.
The reason for this is that the generated congestion starts to prevent the delivery of traffic to the destination. It should be noted that when a packet cannot be sent to its destination on time, all schemes delete that packet.

Figures \ref{fig:random_packets_dropped_3} and \ref{fig:random_packets_dropped_6} depict the packets that are discarded at each node under low traffic load (3 packets per node) and intermediate traffic load (6 packets per node), respectively.
For a fixed traffic load, the original \textit{CGR-DelTime} scheme shows the highest value of dropped packets. This packet loss occurs in each of the 10 nodes, although it is accentuated in nodes 6 to 10, which are those that generate traffic with maximum latency constraints (\textsf{TTL} = 20 s). 
With a lower packet loss, there are the \textit{CGR-MO} schemes with weights 0.25, 0.5 and 0.75. The higher the weight (and importance) given to the number of hops, the better the congestion avoidance capability, and, therefore the lower the packet loss. Finally, \textit{CGR-Hops}, which is equivalent to \textit{CGR-MO} w=1.0, is the scheme with the lowest packet loss and hence the highest delivery ratio. 

\begin{figure}[!b]
\begin{subfigure}{.5\textwidth}
  \centering
  \includegraphics[width=1.0\linewidth]{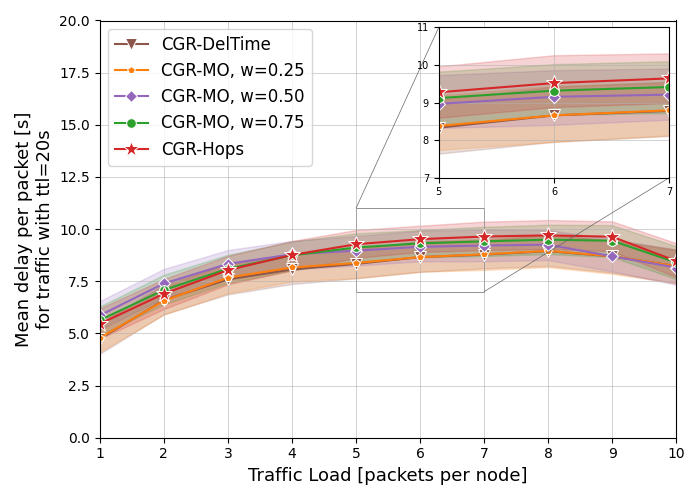} 
  \caption{Mean delay per packet for traffic with \textsf{TTL} = 20 s}
  \label{fig:random_mean_delay_ttl_20}
\end{subfigure}

\vspace{0.2cm}

\begin{subfigure}{.5\textwidth}
  \centering
  \includegraphics[width=1.0\linewidth]{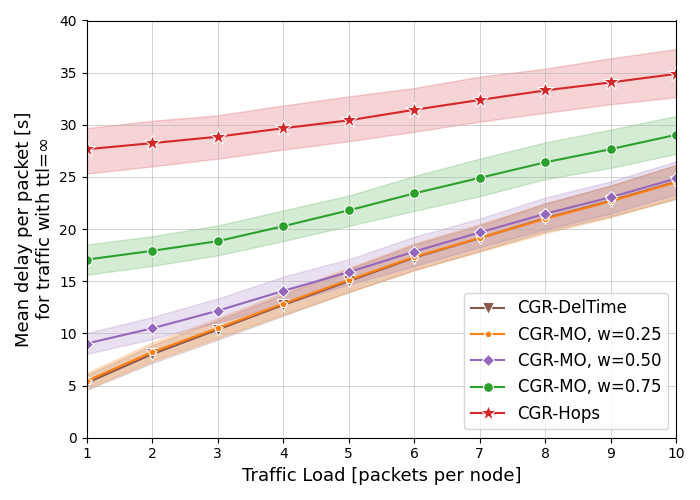} 
  \caption{Mean delay per packet for traffic with \textsf{TTL} $\rightarrow$ $\infty$}
  \label{fig:random_mean_delay_ttl_infinity}
\end{subfigure}
\vspace{0.1cm}
\caption{Simulation results for random networks ($\delta=0.2$).}
\label{fig:random_results_2}
\end{figure}

\begin{figure*}[!h]
\begin{subfigure}{.5\textwidth}
  \centering
  \includegraphics[width=1.0\linewidth]{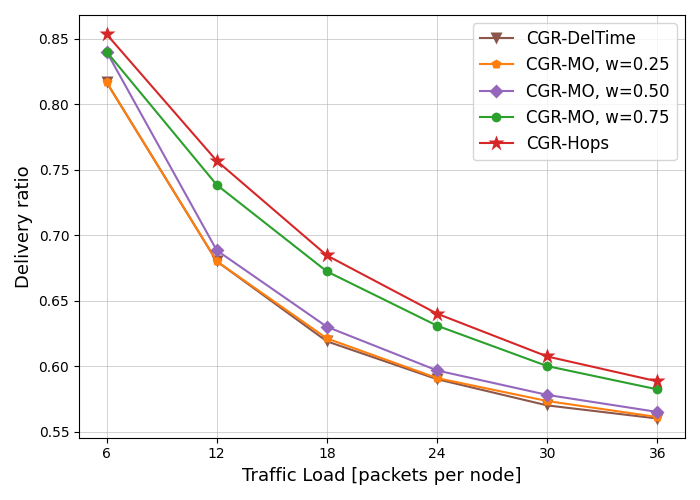}
  \caption{Delivery ratio}
  \label{fig:sub-1}
\end{subfigure}
\vspace{0.4cm}
\begin{subfigure}{.5\textwidth}
  \centering
  \includegraphics[width=1.0\linewidth]{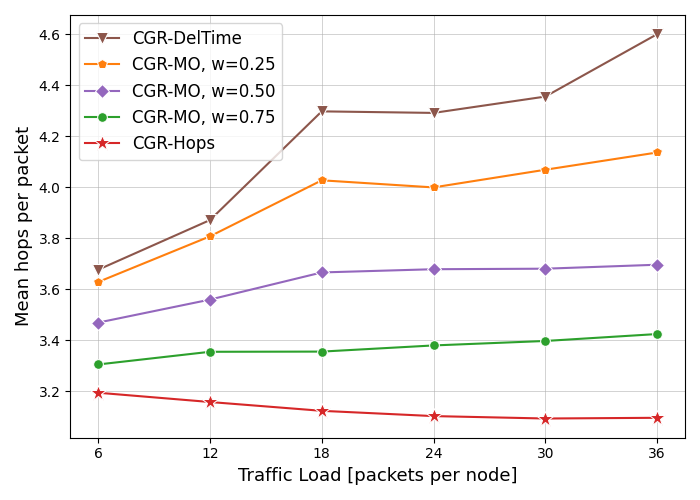}
  \caption{Mean hops per packet}
  \label{fig:sub-2}
\end{subfigure}
\begin{subfigure}{.5\textwidth}
  \centering
  \includegraphics[width=1.0\linewidth]{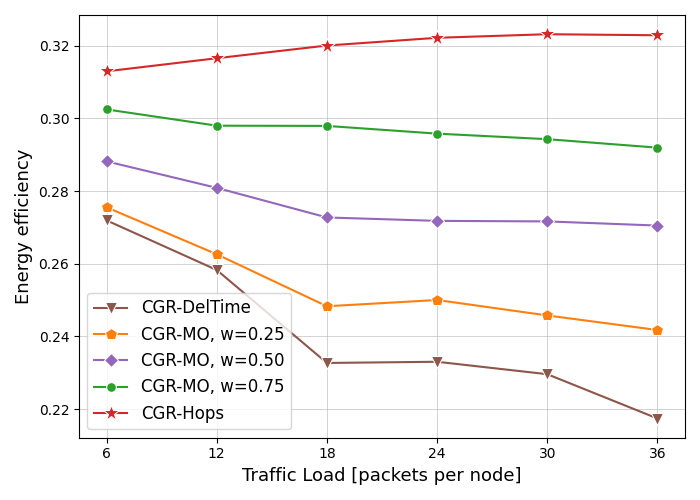} 
  \caption{Energy efficiency}
  \label{fig:sub-3}
\end{subfigure}
\vspace{0.4cm}
\begin{subfigure}{.5\textwidth}
  \centering
  \includegraphics[width=1.0\linewidth]{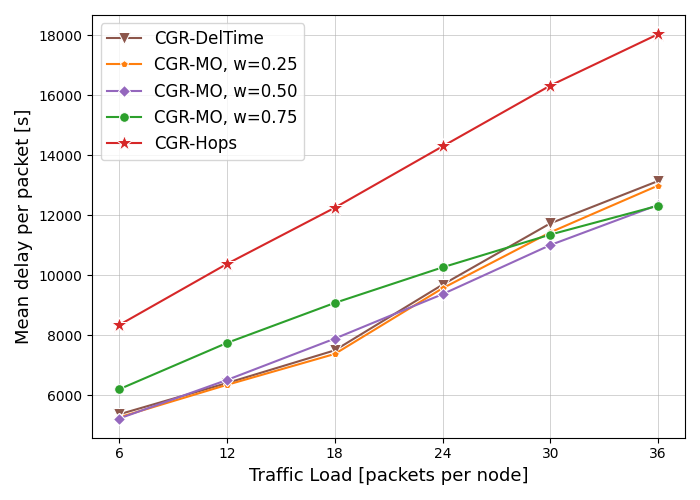} 
  \caption{Mean delay per packet}
  \label{fig:sub-4}
\end{subfigure}
\begin{subfigure}{.5\textwidth}
  \centering
  \includegraphics[width=1.0\linewidth]{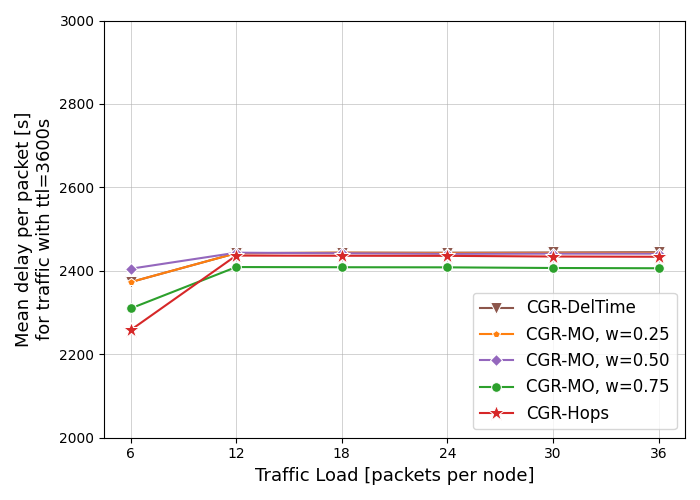}
  \caption{Mean delay per packet for packets with \textsf{TTL} = 3600 s}
  \label{fig:sub-5}
\end{subfigure}
\begin{subfigure}{.5\textwidth}
  \centering
  \includegraphics[width=1.0\linewidth]{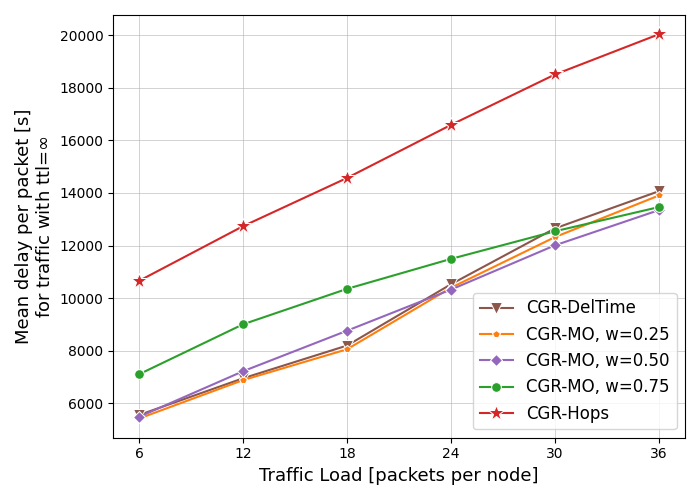}  
  \caption{Mean delay per packet for packets with \textsf{TTL} $\rightarrow$ $\infty$}
  \label{fig:sub-5}
\end{subfigure}
\vspace{0.2cm}
\caption{Simulation results for Walker-Delta constellation.}
\label{fig:walker_results}
\end{figure*}

Regarding the mean hops per packet metric (Figure \ref{fig:random_mean_hops}), \textit{CGR-Hops} obtains the lowest values because it chooses routes with fewer hops. In this case, the \pgm{\textit{ILP model}} results in a higher number of hops per packet because, in addition to satisfying the maximum latency constraint, it aims to deliver packets as quickly as possible, even if that means using more hops. The decreasing trend of hops per packet as the traffic load increases is because congestion forces the model to abandon early states and opt for later states with shorter routes in terms of hops.

In this context, energy efficiency is the measure of the number of successfully delivered packets per transmission. It is the inverse of the mean hops per packet metric and ranges between 0 and 1. Figure \ref{fig:random_energy_efficiency} shows that \textit{CGR-Hops} is the most energy-efficient scheme, with the \pgm{\textit{ILP model}} falling between \textit{CGR-DelTime} and \textit{CGR-Hops}. The model's lack of efficiency can be attributed to its strategy of prioritizing quick delivery, which results in the use of many arcs in early states before moving to later states. In contrast, \textit{CGR-Hops} opts for routes with fewer hops as long as they meet the maximum latency requirement. \textit{CGR-MO} shows values between \textit{CGR-DelTime} and \textit{CGR-Hops} as the weight increases from 0.25 to 0.75, as expected.

Finally, as shown in Figure \ref{fig:random_mean_delay}, the \pgm{\textit{ILP model}} obtains the best metric in terms of delay per packet followed by \textit{CGR-DelTime}, \textit{CGR-MO}, and \textit{CGR-Hops}. However, it is worth noting that this metric is averaging the delay for all generated packets regardless of their maximum latency requirement.  If the mean delay is calculated separately for each type of traffic, we can expect the values to be lower for packets with lower values of \textsf{TTL}. 

Figure \ref{fig:random_results_2} verifies this statement by separating the mean delay for packets with \textsf{TTL} = 20 s, from the mean delay for packets with \textsf{TTL} $\rightarrow$ $\infty$. Figure \ref{fig:random_mean_delay_ttl_20} shows that the curves for all schemes have close values for the entire range of traffic load and that these values are effectively between 0 and 20 seconds. In contrast, Figure \ref{fig:random_mean_delay_ttl_infinity} evidences that the curves have values more distant from each other. The decrease in mean delay when generating 10 packets per node happens because this metric only considers those packets that arrive at the destination, so those packets that are discarded due to congestion do not contribute to increase the metric.

In general, these results reveal that \textit{CGR-Hops} delays traffic that can be delayed (i.e., traffic with higher TTL), thus alleviating congestion and allowing traffic that needs to be delivered earlier to reach its destination on time.

\subsubsection{Walker-Delta Constellation}

The results when the traffic load varies between 6 and 36 packets per source node are plotted in Figure \ref{fig:walker_results}. Since only one Walker-Delta constellation has been simulated, confidence intervals are not plotted. In addition, since the \pgm{\textit{ILP model}} requires extensive run times for scenarios of this size and time, its results are not included here. 

The delivery ratio metric shows that even for low traffic load values, all routing schemes start with a packet loss between 15\% and 20\%. As expected, this loss increases with increasing traffic load and hence network congestion. 
Furthermore, it is observed that the greatest advantage of \textit{CGR-Hops} over \textit{CGR-DelTime} occurs at intermediate traffic load values. When the traffic load is very low, all routing schemes provide solutions that do not lead to competition for the same resources (contacts). When the traffic load is very high, the chances of providing good solutions decrease. 

The mean hops per packet and energy efficiency metrics show behaviors similar to those obtained for the random networks scenario. 
\textit{CGR-Hops} has the highest efficiency as it uses the least number of hops. The \textit{CGR-MO} schemes obtain intermediate and increasing values of energy efficiency as the value of $w$ increases. \textit{CGR-DelTime} leads to the highest number of transmissions to deliver packets to the destination, thus obtaining low energy efficiency. However, a notable difference with the results obtained for random networks is that here the absolute values of the mean number of hops per packet are higher and the energy efficiency values are lower. This is explained by the fact that the simulated Walker-Delta constellation is sparser than the random networks with $\delta=0.2$. Therefore, the routes that packets need to follow to reach their destinations are generally longer in terms of the number of hops.  

Finally, the mean delay per packet metric reveals that \textit{CGR-Hops} pays the cost of obtaining high energy efficiency by generating a higher mean delay per packet. However, to better appreciate which packets are delayed, the average delay per packet is plotted here again, distinguishing according to the \textsf{TTL} configured for each packet. Thus, it can be seen that packets with \textsf{TTL} = 3600 s are delivered to their destination with similar times (and less than 3600 s), while the packets that are effectively more delayed are those that do not have maximum latency restriction (\textsf{TTL} $\rightarrow$ $\infty$). \pgm{A particular effect can be seen in figures \ref{fig:sub-4} and \ref{fig:sub-5} when the traffic load increases to more than 30 packets per node. The curves of \textit{CGR-DelTime} and \textit{CGR-MO, w=0.25} appear to cross the curve of \textit{CGR-MO, w=0.75}. One possible explanation for this is that while schemes such as \textit{CGR-DelTime}, which strictly favor the delay metric, end up exacerbating congestion by trying to get all traffic to the destination sooner, it is also true that these schemes provide greater rerouting opportunities for traffic that is bottlenecked at intermediate nodes. On the other hand, it should be noted here again that since the mean delay per packet metric only considers those packets that actually arrive at their destination, it is not straightforward to compare between schemes when they provide different delivery ratio values.}

\section{Future directions}
\label{sec:futurework}
As future work, we envision further modifications to CGR to improve  performance. In particular considering:
\begin{itemize}
\item \textit{Load balancing}: When all nodes always use the shortest routes, it is likely that multiple routes will share more links, creating a bottleneck in terms of capacity and congestion. To address this issue, we propose different load balancing techniques by implementing round-robin on the $K$ best routes.
\item \textit{Centrality}: In graph theory, the centrality metric measures how many routes pass through a link or node. To avoid congestion, we can use routes through links or nodes that do not have the highest centrality.
\item \textit{Machine learning}: We propose using machine learning models to estimate congestion on a link or node at a given time. We can then use these local estimates to carry out routing and avoid more congested routes but without under-utilizing them.
\item \textit{Uncertain contact plans}: Since contact plans cannot be perfectly implemented in practice, investigating the impact of contact plan uncertainty on traffic routing continues to be a research area that requires special attention.
\end{itemize}

\section{Conclusion}
\label{sec:conclusion}
In this work, we have proposed and evaluated adaptations
to CGR to improve performance when sending traffic with
different latency requirements. 
The results reveal that CGR currently has a limited capability to honor latency constraints. The reason is twofold: on the one hand, it only uses local traffic information. On the other hand, it strictly selects those routes with the lowest estimated delivery time without taking into account the number of hops involved. Since intermediate nodes also generate and carry traffic, this leads to congestion situations that prevent the effective fulfillment of latency requirements. To address these challenges, the proposed schemes explore different ways to consider both the estimated delivery time and the number of hops.  The results show that the new schemes obtain higher delivery rates and better energy efficiency by using fewer hops per packet. In addition, the cost of these benefits is paid for by increasing the delay of those packets without latency constraints. Packets that need to arrive earlier continue to be delivered on time to their respective destinations. This opens the debate on different ways to improve performance without compromising the quality of service required by heterogeneous traffic.

\section*{Acknowledgments}
This work has been supported in part by the National Research Council Canada's (NRC) High Throughput Secure Networks program within the Optical Satellite Communications Consortium Canada (OSC) framework

\balance
\bibliographystyle{IEEEtran}
\bibliography{biblio}

\begin{IEEEbiography}[{\includegraphics[width=1in,height=1.25in,clip]{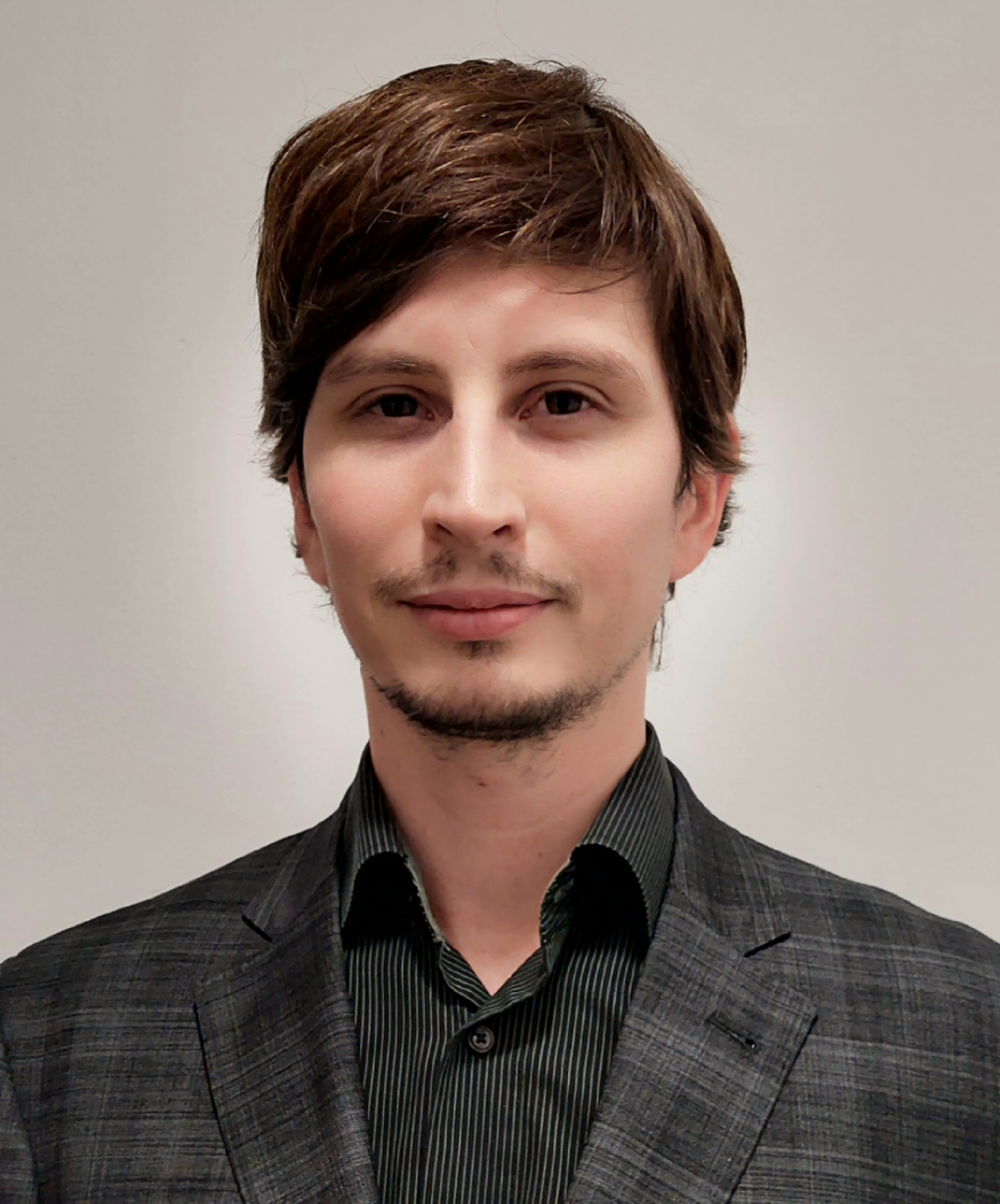}}]{Pablo G. Madoery}
(Graduate Student Member, IEEE) received the B.S. degree in telecommunication engineering from Instituto Universitario Aeronautico, Argentina, in 2015, and the Ph.D. degree in engineering sciences from Universidad Nacional de Cordoba, Argentina, in 2019. He is a postdoctoral fellow at Carleton University on a MITACS fellowship, and an assistant professor of Computer Science at Universidad Nacional de Cordoba, Argentina. He has co-authored 10+ articles in international journals and 20+ in leading conferences in the area of communication protocols and routing algorithms for delay-tolerant satellite networks. Currently, he is researching AI/ML-assisted routing and transport solutions applied to satellite mega-constellations.
\end{IEEEbiography}

\vskip -1\baselineskip plus -1fil

\begin{IEEEbiography}[{\includegraphics[width=1in,height=1.25in,clip]{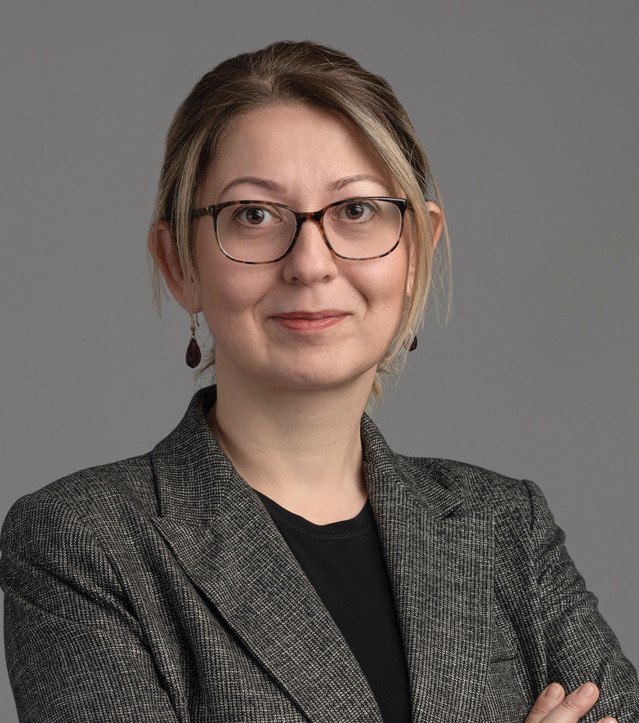}}]{G{\"{u}}ne{\c{s}}~Karabulut~Kurt}
(Senior Member, IEEE) received the B.S. degree with high honors in electronics and electrical engineering from Bogazici University, Istanbul, Turkey, in 2000 and the M.A.Sc. and the Ph.D. degrees in electrical engineering from the University of Ottawa, ON, Canada, in 2002 and 2006, respectively. From 2000 to 2005, she was a Research Assistant with the CASP Group, University of Ottawa. Between 2005 and 2006, she was with TenXc Wireless, Canada. From 2006 to 2008, Dr. Karabulut Kurt was with Edgewater Computer Systems Inc., Canada. From 2008 to 2010, she was with Turkcell Research and Development Applied Research and Technology, Istanbul. Between 2010 and 2021, she was with Istanbul Technical University. She is currently an Associate Professor of Electrical Engineering at Polytechnique Montréal, Montréal, QC, Canada. She is a Marie Curie Fellow and has received the Turkish Academy of Sciences Outstanding Young Scientist (TÜBA-GEBIP) Award in 2019. In addition, she is an adjunct research professor at Carleton University. She is currently serving as an associate technical editor of the \textit{IEEE Communications Magazine}, an associate editor of \textit{IEEE Communication Letters}, an associate editor of \textit{IEEE Wireless Communications Letters}, and an area editor of \textit{IEEE Transactions on Machine Learning in Communications and Networking}. She is a member of the IEEE WCNC Steering Board. She is serving as the secretary of IEEE Satellite and Space Communications Technical Committee and also the chair of the IEEE special interest group entitled "Satellite Mega-constellations: Communications and Networking". She is a Distinguished Lecturer of Vehicular Technology Society Class of 2022.
\end{IEEEbiography}

\begin{IEEEbiography}[{\includegraphics[width=1in,height=1.25in,clip]{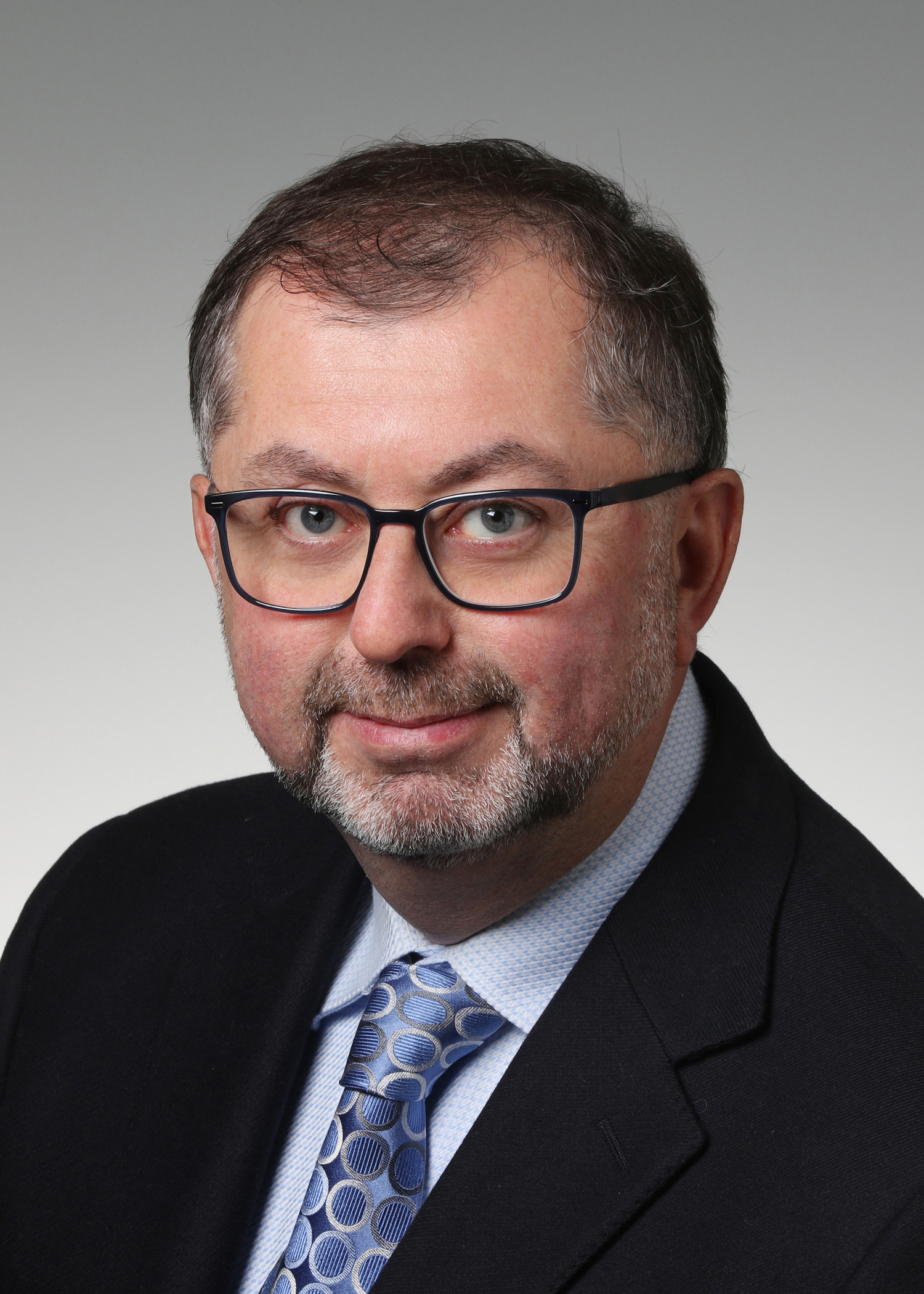}}]{Halim Yanikomeroglu}
(Fellow, IEEE) received the BSc degree in electrical and electronics engineering from the Middle East Technical University, Ankara, Turkey, in 1990, and the MASc degree in electrical engineering (now ECE) and the PhD degree in electrical and computer engineering from the University of Toronto, Canada, in 1992 and 1998, respectively. Since 1998 he has been with the Department of Systems and Computer Engineering at Carleton University, Ottawa, Canada, where he is now a Full Professor. Dr. Yanikomeroglu’s research interests cover many aspects of wireless communications and networks, with a special emphasis on non-terrestrial networks (NTN) in the recent years. He has given 110+ invited seminars, keynotes, panel talks, and tutorials in the last five years. He has supervised or hosted over 150 postgraduate researchers in his lab at Carleton. Dr. Yanikomeroglu’s extensive collaborative research with industry resulted in 39 granted patents. Dr. Yanikomeroglu is a Fellow of the IEEE, the Engineering Institute of Canada (EIC), and the Canadian Academy of Engineering (CAE). He is a Distinguished Speaker for the IEEE Communications Society and the IEEE Vehicular Technology Society, and an Expert Panelist of the Council of Canadian Academies (CCA|CAC). Dr. Yanikomeroglu is currently serving as the Chair of the Steering Committee of IEEE’s flagship wireless event, Wireless Communications and Networking Conference (WCNC). He is also a member of the IEEE ComSoc Governance Council, IEEE ComSoc GIMS, IEEE ComSoc Conference Council, and IEEE PIMRC Steering Committee. He served as the General Chair and Technical Program Chair of several IEEE conferences. He has also served in the editorial boards of various IEEE periodicals. Dr. Yanikomeroglu received several awards for his research, teaching, and service, including the IEEE ComSoc Fred W. Ellersick Prize (2021), IEEE VTS Stuart Meyer Memorial Award (2020), and IEEE ComSoc Wireless Communications TC Recognition Award (2018). He received best paper awards at IEEE Competition on Non-Terrestrial Networks for B5G and 6G in 2022 (grand prize), IEEE ICC 2021, IEEE WISEE 2021 and 2022.
\end{IEEEbiography}

\begin{IEEEbiography}[{\includegraphics[width=1in,height=1.25in,clip]{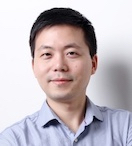}}]{Peng Hu} received his Ph.D. degree in Electrical Engineering from Queen's University, Canada. He is currently a Research Officer with the National Research Council Canada and an Adjunct Professor with the Cheriton School of Computer Science at the University of Waterloo. He has served as an Associate Editor of the IEEE Canadian Journal of Electrical and Computer Engineering, a voting member of the IEEE Sensors Council Standards Committee, and on the organizing/technical committees of industry consortia and international conferences/workshops at IEEE GLOBECOM’23, IEEE ICC'23, IEEE GLOBECOM’21, IEEE PIMRC'17, IEEE AINA'15, etc. His current research interests include satellite-terrestrial integrated networks, autonomous networking, and industrial Internet of Things systems.
\end{IEEEbiography}

\vskip -1.6\baselineskip plus -1fil

\begin{IEEEbiography}[{\includegraphics[width=1in,height=1.25in,clip]{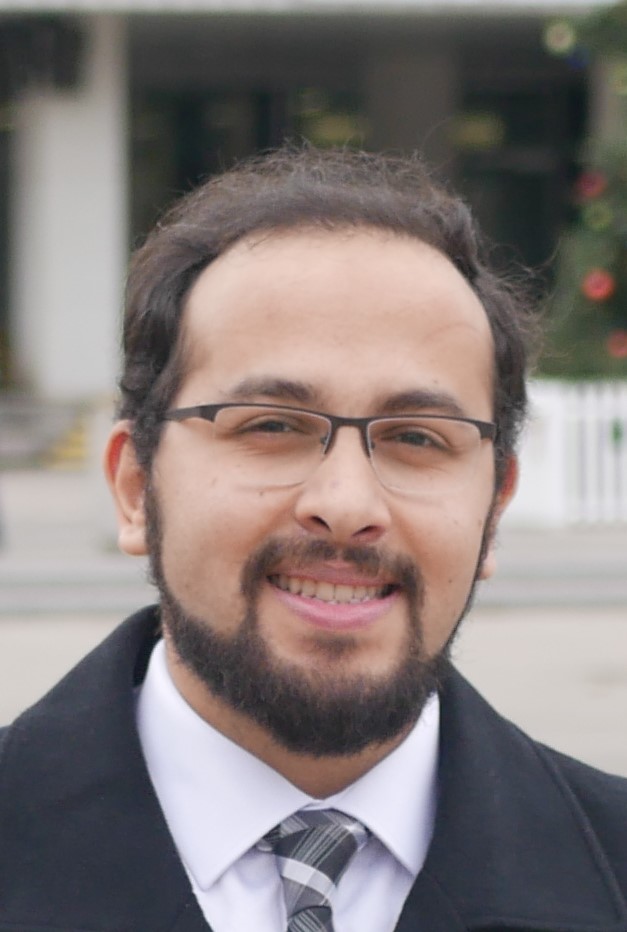}}]{Khaled Ahmed} received his B.S. and MSc degree on wireless communications from Cairo university, Egypt, in 2010 and 2015. He did a PhD and a postdoc in McMaster university, Canada, in 2019 and 2021. His PhD degree was on optical communications, and his postdoc was on machine learning for resource-limited devices. He is currently working as a Member Technical Staff II in the Systems Engineering Department at MDA, Sainte-Anne-de Bellevue, QC, Canada. His current research interests include RF and optical communications for terrestrial and satellite networks, optical beamforming networks, and applied machine learning and deep learning in networks. 
\end{IEEEbiography}

\vskip -1\baselineskip plus -1fil

\begin{IEEEbiography}[{\includegraphics[width=1in,height=1.25in,clip]{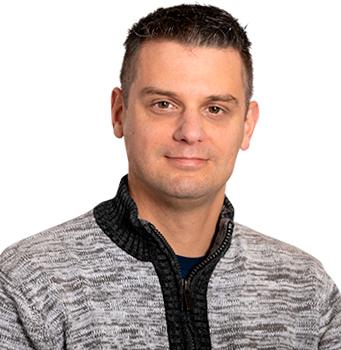}}]{Stéphane Martel} received his Bachelor’s degree in Electrical Engineering from McGill University in 2001. He worked until 2021 in the broadcast television industry where he occupied multiple roles in software development and management. He specializes in network communications and embedded software development.  He joined MDA Satellite systems in 2021 as R\&D product development manager where he contributes to the advancement of space communication technologies.
\end{IEEEbiography}

\vskip -1\baselineskip plus -1fil

\begin{IEEEbiography}[{\includegraphics[width=1in,height=1.25in,clip]{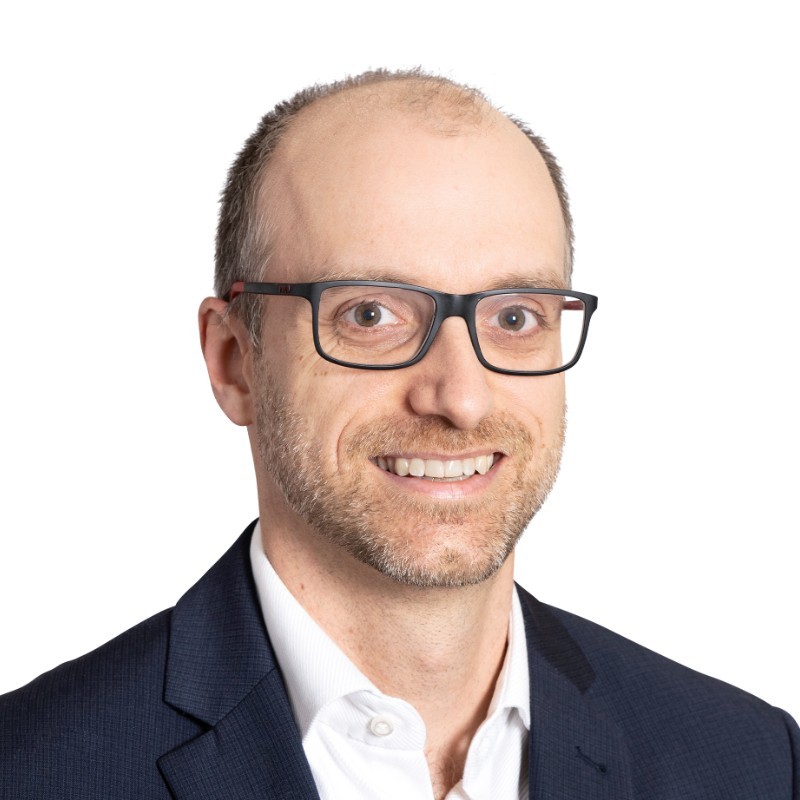}}]{Guillaume Lamontagne}
received the B.Eng. and M.Eng. degrees from the École de Technologie Supérieure, Montréal, QC, Canada, in 2007 and 2009, respectively. His experience in satellite communications started through internships and research activities with the Canadian Space Agency in 2005, and the Centre national d’études spatiales, France, in 2006 and 2008. He joined MDA in 2009 and held various communication systems engineering and management positions before being appointed as the Director of Technology, Payloads in 2019. Through this role, he is leading MDA’s Research and Development activities for satellite communications as well as establishing the related long-term development strategy.
\end{IEEEbiography} 

\end{document}